Toward a better analysis of secreted proteins: the example of the myeloid cells secretome

Mireille Chevallet[1,2], Hélène Diemer[3], Alain Van Dorsselaer[3], Christian Villiers[4], Thierry Rabilloud[1,2]

[1] CEA-DSV/iRTSV/LBBSI, Biophysique et Biochimie des Systèmes Intégrés, CEA-Grenoble, 17 rue des martyrs, F-38054 GRENOBLE CEDEX 9, France

2 CNRS UMR 5092, Biophysique et Biochimie des Systèmes Intégrés, , CEA-Grenoble, 17 rue des martyrs, F-38054 GRENOBLE CEDEX 9, France

3: CNRS UMR7178, Institut Pluridisciplinaire Hubert Curien, ECPM, 25 rue Becquerel, 67087 STRASBOURG Cedex2, France

4: INSERM U823, équipe 8, Immunologie Analytique des pathologies chroniques , Institut Albert Bonniot,

Correspondence :

Thierry Rabilloud, iRTSV/BBSI

CEA-Grenoble, 17 rue des martyrs,

F-38054 GRENOBLE CEDEX 9

Tel (33)-4-38-78-32-12

Fax (33)-4-38-78-44-99

e-mail: Thierry.Rabilloud@ cea.fr

Abstract

The analysis of secreted proteins represents a challenge for current proteomics techniques. Proteins are usually secreted at low concentrations in the culture media, which makes their recovery difficult. In addition, culture media are rich in salts and other compounds interfering with most proteomics techniques, which makes selective precipitation of proteins almost mandatory for a correct subsequent proteomics analysis. Last but not least, the non-secreted proteins liberated in the culture medium upon lysis of a few dead cells heavily contaminate the so-called secreted proteins preparations.
Several techniques have been used in the past for concentration of proteins secreted in culture media. These techniques present several drawbacks, such as coprecipitation of salts or poor yields at low protein concentrations. Improved techniques based on carrier-assisted trichloroacetic acid precipitation are described and discussed in this paper. These techniques have been used to analyse the secretome of myeloid cells (macrophages, dendritic cells) and enabled to analyze proteins secreted at concentrations close to 1 ng/ml, thereby allowing to detect some of the cytokines (TNF, IL-12) secreted by the myeloid cells upon activation by bacterial products.

# 1. Introduction

Proteomics is currently plagued by strong undersampling problems. Owing to the complexity of the proteome and to the huge expression dynamics range of the proteins, all proteomics techniques used to date are far from being comprehensive. Thus, addressing biological problems by proteomics often means to make choices and not to try to tackle the complete proteome at once. One of the possible choices is to reduce the complexity of the proteome to be analyzed, and this has been achieved mainly by using the internal compartimentalization of eukaryotic cells and thus analyzing organellar proteomes (e.g. [1], [2]). Other strategies have also been used, e.g. through the selective isolation of a class of enzymes [3].
Other strategies can however be implemented. When the interest is in cellular communication, it may be interesting to focus the proteomics analysis either in the receiving part of the signal, i.e. membrane receptors. However, membrane proteins are known to be very difficult to analyze by proteomics [4], although recent progress has been made in this area [5], [6]. It may also be interesting to analyze the signalling part itself, i.e. the proteins that are secreted by a given cell type to communicate with other cells. Apart from the analysis of biological fluids, which is a widely developed area in proteomics, only a few examples are present in the literature dealing with the analysis of secreted proteins (e.g. in [7], [8]). This is likely to be due to peculiar problems arising when this analysis is to be made. These problems can be grouped along three axes: i) the low concentration of bona fide secreted proteins, which can go down to the ng/ml range as for the cytokines, ii) the contamination of the authentic secreted proteins by the proteins released in the medium by cell lysis, which occurs even in the best cultures and iii) the contamination by serum proteins, as most animal cell cultures must be carried out in the presence of serum (most often fetal bovine serum), at least in the long term culture range.
In the current work, we have tried to address these three problems, and selected strategies have been devised to achieve a better coverage of the secretome, as exemplified on murine myeloid cells.

2. Material and methods

2.1. Samples

2.1.1. Cell cultures

J774 cells (transformed murine macrophages) were grown in flasks (5% $CO_2$ atmosphere) in DMEM + 5% fetal calf serum up to a density of 0.5 million cells /ml. The medium and detached cells were removed, and the cell layer was washed three times in PBS, and then three times in serum-free DMEM. Serum- and phenol red-free DMEM was added (typically 30 ml for a 175 $cm^2$ flask), and the cells were reincubated at 37°C for 24 to 48 hours.
The cell supernates were then aspirated, collected ,and centrifuged at 1000g for 5 minutes (4°C) to pellet detached cells and large debris. The supernatant was collected and then centrifuged for 1 hour at 100,000g (4°C) to pellet smaller debris and vesicles. The final supernatant was saved and kept frozen (-80°C) until use.
The cell layer remaining in the flask after the incubation period in serum-free medium was scraped in a few ml of serum-free and phenol red-free DMEM. It was washed 3 times in PBS and resuspended in 10 times the pellet volume of serum-free and phenol red-free DMEM. This suspension was further lysed by 10 strokes of a Dounce homogeneizer at 4°C. The lysate was clarified by centrifugation at 1000g for 5 minutes (4°C). The supernatant was collected and then centrifuged for 1 hour at 100,000g (4°C) to pellet smaller debris, organelles and vesicles. This final supernatant, called cytosolic extract, was saved and kept frozen (-80°C) until use

Alternatively, complete cell extracts were prepared by direct lysis of a cell suspension in 7M urea, 2M thiourea, 4% CHAPS 10mM spermine base and 5mM tris(carboxyethyl) phosphine (final concentrations). After 1 hour at room temperature, the nucleic acids were removed by ultracentrifugation (200,000g 1 hour, room temperature), and the supernatant was used for proteomics experiments.

Murine dendritic cells were produced from bone marrow progenitors as described previously [9]. The cells were washed and transferred to serum-free medium(1 million cells /ml) as described for the J774 cells, and the dendritic cells were activated by the addition of 1 μg of E. coli lipopolysaccharide per ml of culture medium for 24 hours. The resulting medium and cell layer were then treated and stored as described above.

All experiments were performed at least three times arising from different culture batches.

**2.2. Conditioned medium treatment**

The proteins present in the conditioned media were concentrated by various procedures. In some cases, the protein concentration before and after concentration was assayed by a dye-binding assay (Rio-Rad)

2.2.1. Ultrafiltration
The conditioned media were centrifuged on a Millipore Ultrafree15 device at 4000g for 2 hours. This long time was required because of membrane clogging. The resulting concentrate was diluted 4 times in 1.25x strip rehydration buffer and applied to the IPG strips.

2.2.2. Adsorption to diatomaceous earth [10]
The conditioned medium (20 ml) was agitated on a rotating wheel for 1 hour at 4°C with 20 mg of acid-washed diatomaceous earth (from Sigma). The suspension was then centrifuged for 5 minutes at 1000g (4°C) and the supernatant was drawn off. The diatomaceous earth pellet was then resuspended in 1 ml of Tris-HCl 5mM pH 7.5 and transferred in a 1.5ml centrifuge tube. The centrifugation step was repeated once. The final pellet was resuspended in 0.4 ml of extraction solution (7M urea, 2M thiourea, 4% CHAPS and 5mM tris(carboxyethyl) phosphine) and agitated for one hour (rotating wheel) at room temperature. The suspension was centrifuged at 2000g for 5 minutes (room temperature) and the supernatant was collected and used directly for 2D electrophoretic separation.

2.2.3. Adsorption to hydrophobic resin [11]
The process was very similar to the one described above, except that diatomaceous earth was replaced by the Strataclean ® resin.

2.2.4. Phenol extraction [12]

20 ml of conditioned medium was mixed with 2.3 ml of water-saturated phenol. The extraction was carried out at room temperature for 15 minutes with occasional vortexing. The phases were then separated by centrifugation (5 minutes, 2000g, room temperature). The upper water phase was removed, and the phenol phase was transferred to a 2ml microcentrifuge tube. 1 μl of 20% SDS and 1 ml of diisopropyl ether were then added. After mixing by vortexing, the emulsion was centrifuged at 10,000g for 10 minutes. The upper organic phase was discarded, and the lower water phase was washed again with 1 ml of ether. After centrifugation as described above, the lower water phase was collected, and diluted in concentrated denaturing solution to end with a standard extraction solution (7M urea, 2M thiourea, 4% CHAPS and 5mM tris(carboxyethyl) phosphine).

2.2.5. TCA precipitation

This protocol was derived from the DOC-TCA protocol previously described [13]. 20 ml of conditioned medium were first cooled on ice in a high speed centrifuge tube. Sodium deoxycholate or sodium lauroyl sarcosinate (NLS) were added to a final concentration varying from 0.01 to 0.1%. After mixing, TCA was added to a final 7.5% concentration, and the solution was precipitated on ice for 2 hours. The mixed protein-detergent precipitate was collected by centrifugation (10,000g, 10 minutes 4°C). The supernatant was carefully removed, 2 ml of tetrahydrofuran (precooled in ice) were added to the pellet and vortexing was carried out until the pellet unstuck from the bottom of the tube, and dissolved almost completely. Centrifugation was carried out as described above. The supernatant was removed, and the nearly invisible pellet was washed again with 2ml of THF. Finally, the pellet was redissolved in 0.4 ml extraction solution with the help of a sonicator bath (30 minutes extraction).

**2.3. Electrophoresis**

2.3.1. IEF

Home made 160mm long 4-8 or 3-10.5 linear pH gradient [14] gels were cast according to published procedures [15]. Four mm-wide strips were cut, and rehydrated overnight with the sample, diluted in a final volume of 0.6 ml of rehydration solution (7M urea, 2M thiourea, 4% CHAPS, 04% carrier ampholytes (Pharmalytes 3-10) and 100mM dithiodiethanol [16], [17]). The strips were then placed in a multiphor plate, and IEF was carried out with the following electrical parameters

100V for 1 hour, then 300V for 3 hours, then 1000V for 1 hour, then 3400 V up to 60-70 kVh.

After IEF, the gels were equilibrated for 20 minutes in Tris 125mM, HCl 100mM, SDS 2.5%, glycerol 30% and urea 6M. They were then transferred on top of the SDS gels and sealed in place with 1% agarose dissolved in Tris 125mM, HCl 100mM, SDS 0.4% and 0.005% (w/v) bromophenol blue.

2.3.2. SDS electrophoresis and protein detection

10%T gels (160x200x1.5 mm) were used for protein separation. the gel buffer system is the classical Laemmli buffer pH 8.8, used at a ionic strength of 0.1 instead of the classical 0.0625.

The electrode buffer is Tris 50mM, glycine 200mM, SDS 0.1%.

The gels were run at 25V for 1hour, then 100V until the dye front has reached the bottom of the gel. Detection was carried out by ammoniacal silver staining [18]. Gel image analysis was carried out with the Melanie2 software package.
The spots of interest were excised by a scalpel blade and transferred to a 96 well microtitration plate. Destaining of the spots was carried out by the ferricyanide-thiosulfate method [19] on the same day than silver staining to improve sequence coverage in the mass spectrometry analysis [20]

**2.4. Mass spectrometry**

2.4.1. In gel digestion

In gel digestion was performed with an automated protein digestion system, MassPrep Station (Waters, Manchester, UK). The gel plugs were washed twice with 50 μL of 25 mM ammonium hydrogen carbonate ($NH_4HCO_3$) and 50 μL of acetonitrile. The cysteine residue were reduced by 50 μL of 10 mM dithiothreitol at 57°C and alkylated by 50 μL of 55 mM iodoacetamide. After dehydration with acetonitrile, the proteins were cleaved in gel with 10 μL of 12.5 ng/μL of modified porcine trypsin (Promega, Madison, WI, USA) in 25 mM $NH_4HCO_3$. The digestion was performed overnight at room temperature. The generated peptides were extracted with 60% acetonitrile in 5% acid formic.

2.4.2. MALDI-MS

MALDI-TOF mass measurements were carried out on UltraflexTM TOF/TOF (Bruker DaltonikGmbH, Bremen, Germany). This instrument was used at a maximum accelerating potential of 25kV in positive mode and was operated in reflectron mode. The sample were prepared by standard dried droplet preparation on stainless steel MALDI targets using α-cyano-4-hydroxycinnamic acid as matrix.
The external calibration of MALDI mass spectra was carried out using singly charged monoisotopic peaks of a mixture of bradykinin 1-7 (m/z=757.400), human angiotensin II (m/z=1046.542), human angiotensin I (m/z=1296.685), substance P (m/z=1347.735), bombesin (m/z=1619.822), renin (m/z=1758.933), ACTH 1-17 (m/z=2093.087) and ACTH 18-39 (m/z=2465.199). To achieve mass accuracy, internal calibration was performed with tryptic peptides coming from autolysis of trypsin, with respectively monoisotopic masses at m/z = 842.510, m/z = 1045.564 and m/z = 2211.105 . Monoisotopic peptide masses were automatically annotated using Flexanalysis 2.0 software.

2.4.3. NanoLC-MS/MS

Nano-LC-MS/MS analysis was performed either using a CapLC capillary LC system (Waters), coupled to a hybrid Quadrupole Time-Of-Flight mass spectrometer (Q-TOF II, Waters).
From each sample, 6.4 μL was loaded on a precolumn, before chromatographic separation on a C18 column (LC Packings C18, 75 mm id, 150 mm length). The gradient was generated by the CapLC at a flow rate of 200 nL/min. The gradient profile consisted of a linear one from 90% of a water solution acidified by 0.1% HCOOH vol/vol (solution A), to 40% of a solution of CH3CN acidified by 0.1% HCOOH vol/vol (solution B) in 30 min, followed by a second gradient ramp to 75% of B in 1 min. Data acquisition was piloted by MassLynx software V4.0. Calibration was performed using adducts of 0.1% phosphoric acid (Acros, NJ, USA) with a scan range from m/z 50 to 1800. Automatic switching between MS and MS/MS modes was used. The internal parameters of Q-TOF II were set as follows. The electrospray capillary voltage was set to 3.5 kV, the cone voltage set to 35 V, and the source temperature set to 90°C. The MS survey scan was m/z 300–1500 with a scan time of 1 s and an interscan time of 0.1 s. When the peak intensity rose above a threshold of 15 counts/s, tandem mass spectra were acquired. Normalized collision energies for peptide fragmentation were set using the charge-state recognition files for 1+, 2+, and 3+ peptide ions. The scan range for MS/MS acquisition was from m/z 50 to 2000 with a scan time of 1 s and an interscan time of 0.1 s. Fragmentation was performed using argon as the collision gas and with a collision energy profile optimized for various mass ranges of precursor ions. Data processing was done automatically with the ProteinLynx Process (Micromass) module.

2.4.4. MS and MS/MS Data analysis

The research in SwissProt and TrEMBL was done using the programm MASCOT [21] (Matrix Science Ltd., London UK). The research was carried out in all species. One missed cleavage per peptide was allowed and the following variable modifications carbamidomethylation for cysteine and oxidation for methionine were taken into account. For Peptide Mass Fingerprint, a mass tolerance of 50 ppm was allowed and for MS/MS ion search a mass tolerance of 0.25 Da.

Immunoassay of cytokines
The cytokines were measured in the cell supernatants by ELISA using the OptEIA set for mouse cytokines from Pharmingen according to the procedures recommended by the manufacturer

## 3. Results

Various protein concentration methods were evaluated for their performance and compatibility with 2D electrophoresis. Figure 1 describes the results obtained by ultrafiltration, diatomaceous earth and hydrophobic resin. Various problems were encountered with these methods, either poor recovery or loss of resolution in the IEF dimension and therefore in 2D electrophoresis. Better results were obtained by using phenol extraction and TCA precipitation (figure 2). In the latter method, lauroyl sarcosinate performed slightly better than deoxycholate, superior itself to phenol extraction. To investigate the apparent threshold effects in the two precipitation protocols, we performed a comparative experiment with cytosolic extracts diluted to various levels and processed according to the TCA/deoxycholate and the TCA/lauroylsarcosinate protocols. The results are shown on figure 3, and demonstrate that the TCA/lauroylsarcosinate method is clearly superior. In these experiments, image comparison between the unprecipitated samples and the ones precipitated with the TCA/lauroylsarcosinate method provided a precipitation yield estimation, close to 75%.In addition, this experimental setup showed that the protein patterns were not distorted by the protein precipitation, showing that no strong counterselection of proteins was operated by the concentration method chosen. However, it cannot be ruled out that some proteins are missed by this concentration method, but this is true for any concentration methods. In addition, repeated experiments demonstrated that a lauroylsarcosinate solution no older than 2 weeks was needed for optimal results (data not shown). The TCA/lauroylsarcosinate method was then selected for secretome studies. The first of these studies was carried out on the J774 murine macrophage cell line. The 75% yield figure was confirmed at this stage by comparing the protein amount present in 20 ml of conditioned medium (140-150 μg) and the amount recovered after precipitation and solubilization and loaded onto the gels (100-110 μg), giving a 78% yield.

A typical gel with the corresponding identifications obtained by mass spectrometry is shown on figure 4 and Table 1. The spots were randomly selected, except that highly crowded areas were avoided (as they give difficult to assign, multiple identifications) as well as minor spots. From these data, it clearly appeared that contamination of authentic secreted proteins by cytosolic proteins released in the culture medium upon cell lysis was a major issue to be addressed. This contamination is linked to various classes of proteins, including cytoskeletal proteins (e.g. cofilin, moesin, tropomyosin, not even, to quote actin) and major metabolic enzymes (e.g. aldolase, phosphoglycerate kinase, phosphogluconate dehydrogenase, lactate dehydrogenase). Even with extensive washing of the cells, bovine serum proteins could still be found (e.g. albumin, alpha 2S glycoprotein). In order to cope with this problem, we tried to find the best cell extract to be used as a control. These trials were made on another type of myeloid cells, namely dendritic cells. As a matter of facts, a classical extract, made from direct lysis of the cells in denaturing medium, is not the best control, as exemplified on figure 5. An extract made by lysis of cells in culture medium was much more comparable to the 2D pattern obtained from conditioned media. In addition, such an extract can easily be treated by the same procedures than conditioned media to take into account biases introduced by the protein concentration methods.

This approach was used to study the secretome of dendritic cells, as shown on figure 6. Comparative analysis of conditioned media obtained from dendritic cells, either immature or activated with bacterial lipopolysaccharide, vs cytosolic extract arising from the same cells was carried out. At equal conditioned media volume, the gels corresponding to immature dendritic cells (figure 6A) contained much less proteins than those corresponding to LPS-activated dendritic cells (figure 6B). This is linked both to the secretion of proteins (especially those linked to inflammatory processes) by activated dendritic cells, but also to the increased cell mortality induced by LPS [22] [23].

Spots obviously present in higher concentrations (as derived from gel analysis data) in the cytosolic extract (shown of figure 6C) were excluded from the excision-identification process, and priority was given to those present in higher concentrations in conditioned media, i.e. spots exhibiting a two-fold enrichment in cell conditioned media, compared to cytosolic extracts, as assessed by gel image. The identification results are shown in Table 2. Although contaminating cytosolic proteins were still present in the identification list, most of the proteins are either bona fide, well known secreted proteins or lysosomal proteins (mainly cathepsins, which have been found in conditioned media in other studies [24]). Among the secreted proteins, we could identify two cytokines, IL12 and TNF, (see mass spectrometry spectra on figure 7) which are known to be produced by dendritic cells at concentrations in the low nanogram/ml range [9]. The values and the general cytokinic profile were verified by immunoassay on the supernatants used for the proteomics analysis ( figure 8).

## 4. Discussion

A variety of methods have been described to achieve protein concentration from dilute extracts such as culture media. From the data already present in the literature, we ruled out two classes of methods. The first excluded class was precipitation with organic solvents (ethanol and acetone being most often used). A minimal sevenfold volume excess of solvent over aqueous phase must be used, and these methods are therefore not easily adaptable to the analysis of large volumes of conditioned medium. This drawback was quite important, as large quantities of samples must be analyzed if low abundance proteins (down to ng/ml) are to be detected.

The second excluded class was the dye precipitation method [25] [26]. This method was excluded because it selects against glycoproteins [27], which are a major class of secreted proteins. Among the selected methods, the hydrophobic resin adsorption method also counterselected hydrophilic proteins. In addition, adsorption of large quantities of conditioned media gave rise to important 2D electrophoresis artefacts. These artefacts seemed to be dependent on the resin batches and on the amount of media treated. Decent results could be obtained for 10 ml of medium but not for 20 ml (data not shown). This lack of robustness prompted us not to select this method.

This combination of lack of efficiency and electrophoretic artefacts was also found with ultrafiltration. Although different results might be obtained with various devices, e.g. due to different membrane performance, our results did not prompt us to investigate this method much further. In addition, leakage of low molecular weight proteins is expected.

Protein partition in phenol appeared as an alternative deserving some studies. Phenol is known to be able to extract proteins very efficiently, and to allow an important concentration factor [28]. The problem is in fact not to extract proteins in the phenol phase, but to recover them from this phase. A few protocols are available to achieve this task, the most popular being the ammonium acetate/methanol precipitation [29] and the double phase extraction [12]. The former protocol was quite erratic in our hands on our dilute extracts, while the latter performed better. The basis of this protocol is to induce a phase separation in the phenol phase with an hydrophobic solvent. This solvent will solubilize the phenol but phase out the water contained in the phenol phase of a water/phenol extraction, and the proteins are also repelled from this hydrophobic organic phase and end up either as a precipitate on the wall of the tube or in the water phase. This is why we added a small amount of SDS just prior to this final phase separation, both to obtain a minimal amount of water phase and to try to keep the proteins soluble in the water phase.

Albeit encouraging, this method did not perform as well as the carrier-assisted TCA precipitation. Compared to other descriptions of the latter method (e.g. in [30] ) the use of 2D electrophoresis impose to remove most of the carrier and remaining acid from the initial pellet, which may explain the relatively poor results obtained with this method in our case. This acid and detergent removal was achieved by the use of ether type solvents, which are both better solvents for carrier and TCA than acetone and less efficient solvents for proteins than alcohols. Among the ether-type solvents, tetrahydrofuran was selected because it is less volatile than ether, and thus easier to handle, but also more volatile than dioxane, and thus easier to remove from the final urea-thiourea extract. In addition, the initial carrier-protein pellet still contains some water, and a liquid film along the walls of the centrifuge tube is occasionally present. The use of diethylether will result in a two-phase system leading to poor removal of the carrier. Oppositely, tetrahydrofuran is able to absorb some water, so that a single phase is obtained, which is more efficient to remove the carrier molecule and the remaining water.

In order to optimize the choice of the carrier molecule, we tested several carboxylate surfactants. These molecules are water soluble at neutral pH, but become water-insoluble and solvent-soluble at acidic pH, when the carboxylic acid becomes protonated. We reasoned that surfactants able to bind more avidly to proteins than bile salts should be better carriers, so that we tested surfactants with linear alkyl tails. Among those, fatty acids salts (e.g. sodium laurate) prove unsuitable, as the carboxylic acid is of low density and thus floats on top of the water phase upon centrifugation. Sodium lauroyl sarcosinate and sodium lauryl oligooxyethylene glycolate proved equally efficient and more efficient than deoxycholate. Lauroyl sarcosinate was selected for its chemical homogeneity and purity.

In this carrier-assisted TCA precipitation protocol, several caveats exist, especially when relatively large volumes of samples are processed. Efficient protein precipitation requires high speed centrifugation (typically 10,000g), and the presence of the carrier induces a rather large pellet. However, this pellets vanishes in the organic solvent, but it was found that it was not efficient to change the tube at this stage, although it was very tempting for the final extraction in a low volume of sample solution. We believe that the proteins in ether are very insoluble, so that they are present either on the tube wall or will stick to the pipetting device. We also tried a double precipitation protocol, in which the initial carrier-protein pellet is dissolved in a basic medium [31], transferred into a small tube, and then precipitated again with TCA and finally washed with tetrahydrofuran. Severe losses as well as charge heterogeneity artefacts were encountered with this protocol (data not shown).

Despite these caveats, the sarcosinate-TCA protocol proved to be efficient, as shown by the ability to display and identify proteins expressed in the low ng/ml range such as TNF or IL12. Due to the yield of the 2D process itself [32], this however required processing of large sample volumes.

Apart from the protein concentration issues, the analysis of secreted proteins is plagued by contamination problems arising from the serum used for the cultures and from the intracellular proteins released in the medium upon cell lysis. In some cases (e.g. in [7] ) this contamination obscures almost completely the authentic secreted protein profile. Some serum proteins are still present in conditioned media even after thorough washing of the culture vessels and cell layer. However, the use of heterologous serum (e.g.bovine serum for murine cells) allows most of the time to discriminate between the authentic secreted proteins and the serum proteins at the mass spectrometry stage. At least in the 2D gel-based analyses, the sequence coverage for each protein is generally sufficient to get species-specific peptides enabling this discrimination. The situation is far less positive for lysis-derived contaminants, which are even more prominent in serum-free cultures because this reduces cell viability. This problem is underlined by the comparison between macrophages, immature and activated dendritic cells mad in our work. While immature dendritic cells have a very low percentage of dead cells (1% or less, activated dendritic cells and cultured macrophages go up to ca 5% mortality, and this is the main cause for the completely different patterns obtained. This also shows that comparison of secretomes is not always the best way to provide adequate controls. Our suggestion is to use a comparative approach between the conditioned medium and a cytosolic extract, but better approaches are still to be introduced, apart from those using autoradiography of labeled proteins [33].

In conclusion, the protocols we have introduced here, namely the efficient protein precipitation from large sample volumes and the use of a sample comparison process, allowed us to detect even low-abundance secreted proteins (in the low ng/ml range). However, the overall process is still not perfect, as there are some cytokines (e.g. IL1 or IL6) that we could not detect. In addition,, the lauroyl sarcosinate-TCA protein concentration protocol introduced here is not restricted to 2D gel-based proteomics. As this protocol goes through a protein pellet stage, this protein pellet should be dissolvable in other protein denaturing solvents used for other types of proteomics analyses, thereby complementing other approaches [34].

Acknowledgments.

Anne Marie Laharie is thanked for performing the cytokine immunoassays, Anne Papaioannou for expert assistance in cell culture, and Sylvie Luche is thanked for her expert work in proteomics. Personal support from the CNRS to TR is also acknowledged, as well as a grant from the Région Rhone Alpes (analytical chemistry subcall, priority research areas call 2003-2006)

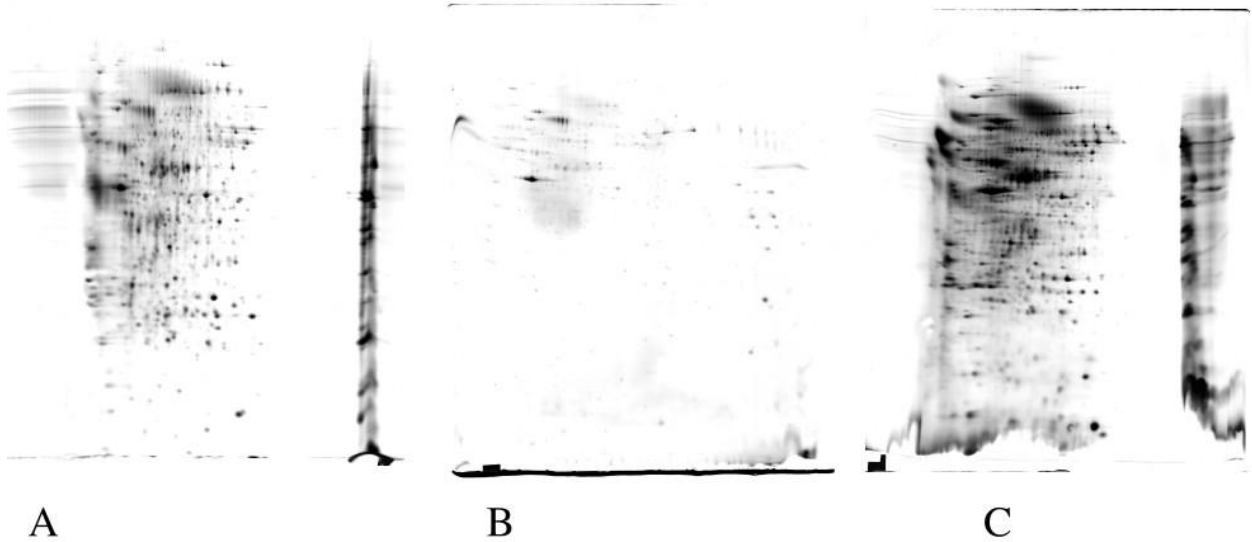

Figure 1: comparative results obtained with filtration/adsorption methods
12ml of medium conditioned for 24 hours by J774 macrophage cells were concentrated by the following methods:
A: ultrafiltration; B: adsorption onto diatomaceous earth; C: adsorption onto a hydrophobic resin. The recovered proteins were analyzed by 2D gel electrophoresis (pH gradient linear 3-10.5) second dimension 10% acrylamide gels, and detected by silver staining. The artifacts likely caused by ionic compounds can be easily seen on gels A and C, while the low overall yield is obvious in gels A and B

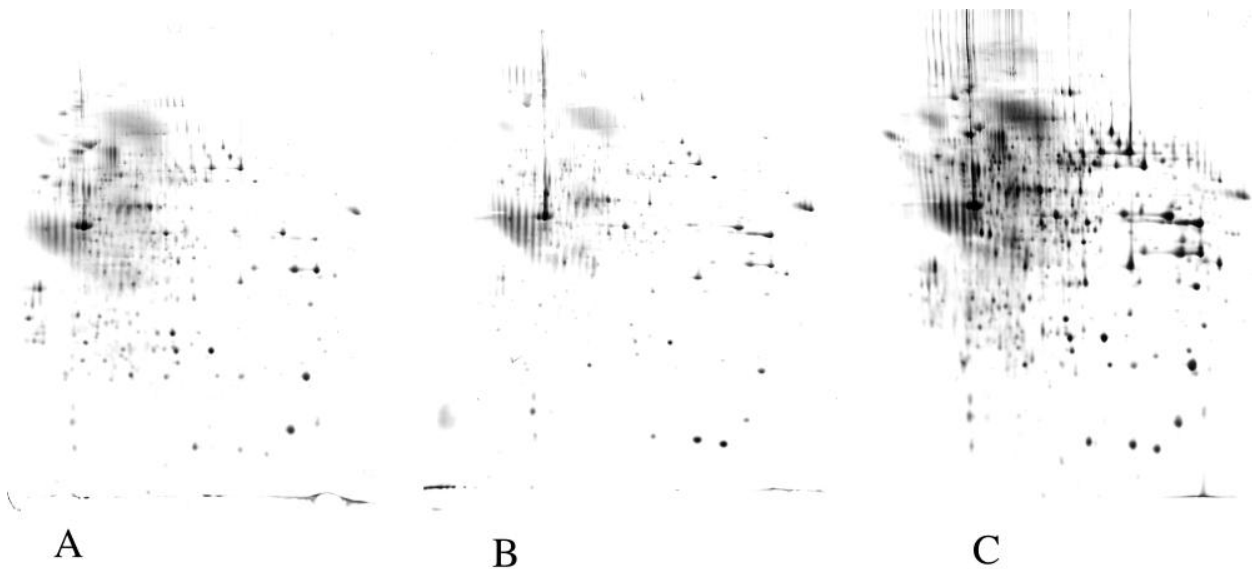

Figure 2: comparative results obtained with partition /precipitation methods
12ml of medium conditioned for 24 hours by J774 macrophage cells were concentrated by the following methods:
A: phenol extraction; B: TCA-DOC precipitation; C: TCA-lauroylsarcosinate (NLS) precipitation. The proteins were analyzed as described in figure 1. The increase in yield with the TCA-NLS method is obvious.

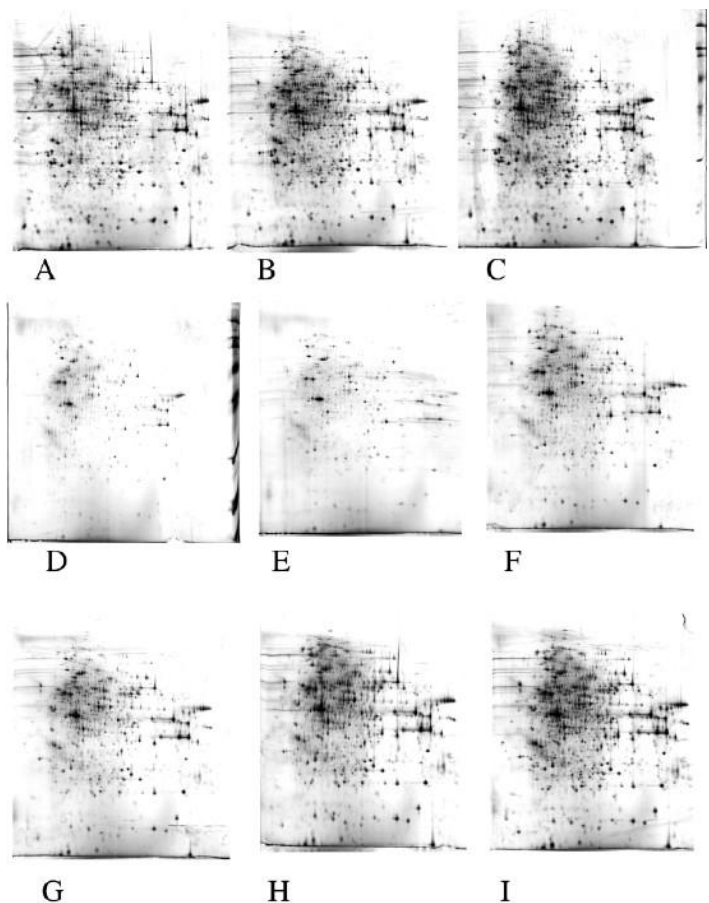

Figure 3: estimation of protein precipitation efficiency
Various amounts of cytosolic extracts from J774 cells were analyzed directly by two-dimensional electrophoresis (panels A to C) or diluted to 20 ml in culture medium and precipitated by the TCA/deoxycholate method (panels D to F) or the TCA/ lauroylsarcosinate method (panels G to I). Panels A, D, G: 50μg of proteins analyzed. Panels B,E,H: 75μg of proteins analyzed. Panels C,F,I: 100μg of proteins analyzed. All other electrophoretic conditions as described in figure 1. The precipitation yield was estimated by comparing the gels obtained with the precipitation methods with the control gels.

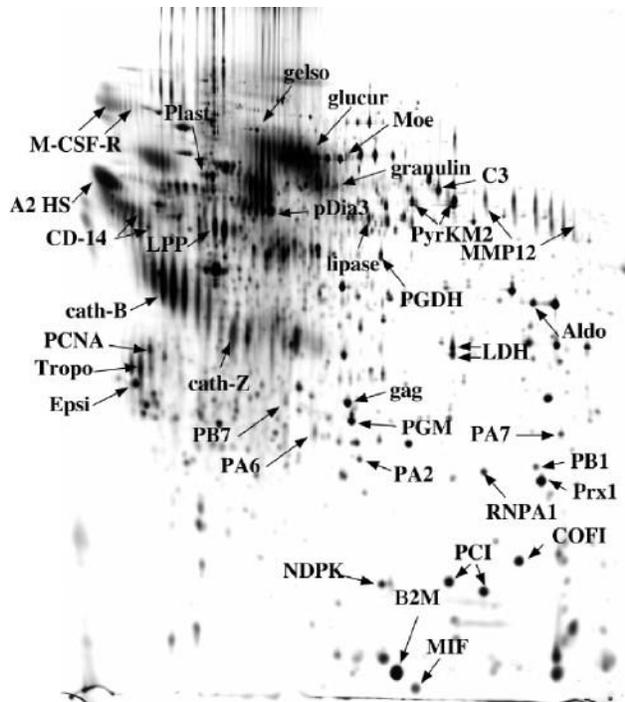

Figure 4: Analysis of secreted proteins produced by the J774 macrophage cell line
20 ml of medium conditioned for 24 hours by J774 macrophage cells (i.e. 10 million cells)were concentrated by the TCA-NLS method. The proteins were analyzed and detected as described in figure 1. Protein spots were excised, destained and proteins were identified through mass spectrometry analysis (see Table 1).

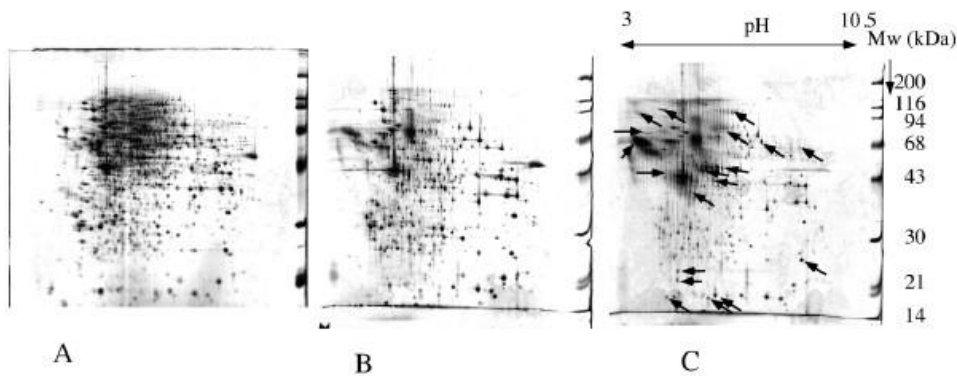

Figure 5: compared analysis of total extract, cytosolic extract and secreted proteins.
12 ml of medium conditioned for 24 hours by dendritic cells were concentrated by the TCA-NLS method. The amount of proteins was evaluated to 150μg. equal amounts of cytosolic extract and complete cell extract in urea-thiourea were analyzed for comparative purposes. All sseparation and detection procedures as in figure 1.
A: complete cell extract; B: cytosolic extract; C: secreted proteins
This figure shows that a complete cell extract may be a less relevant control for comparative analysis than a cytosolic extract. With this latter extract, much less proteins appear to be increased in the secreted fraction (see arrows)

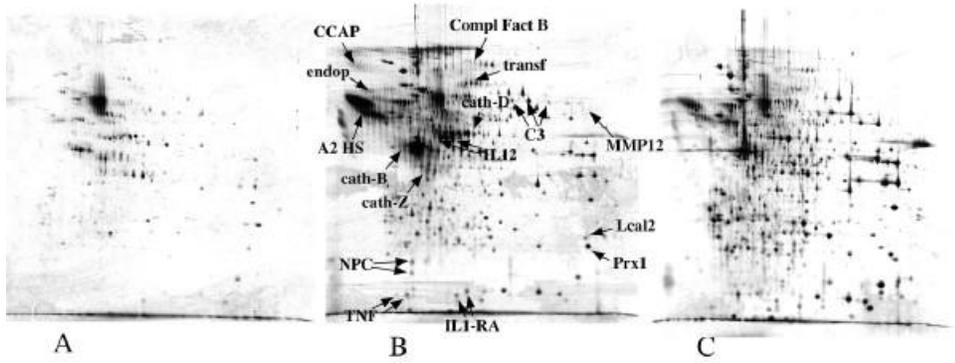

Figure 6: Analysis of secreted proteins produced by the murine activated dendritic cells. 20 ml of medium conditioned for 24 hours by immature dendritic cells (panel A) LPS-activated dendritic cells (panel B), 20 million cells in each case, were concentrated by the TCA-NLS method a,d compared to 100µg of prteoins derived from a cytosolic extract (panel C) . The proteins were analyzed and detected as described in figure 1. Protein spots were excised, destained and proteins were identified through mass spectrometry analysis (see Table 2).

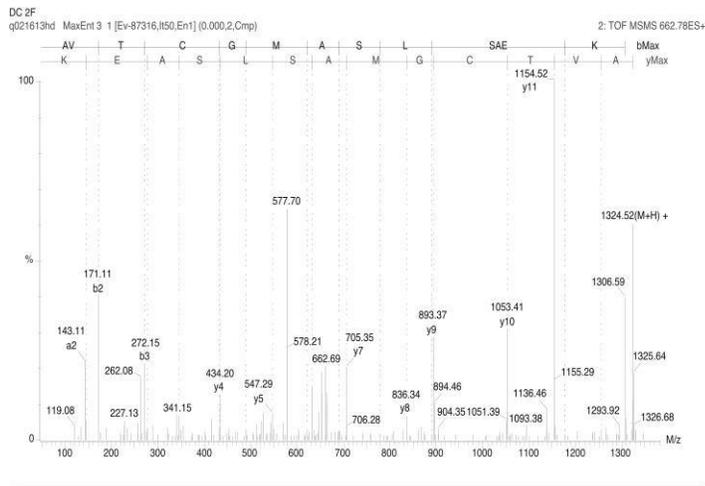

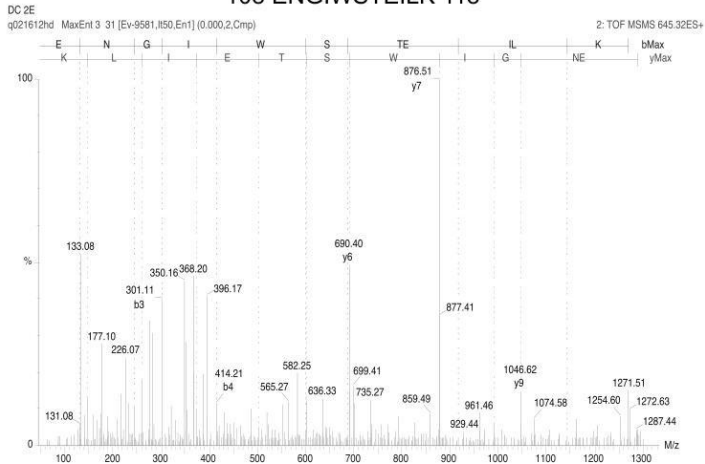

7A

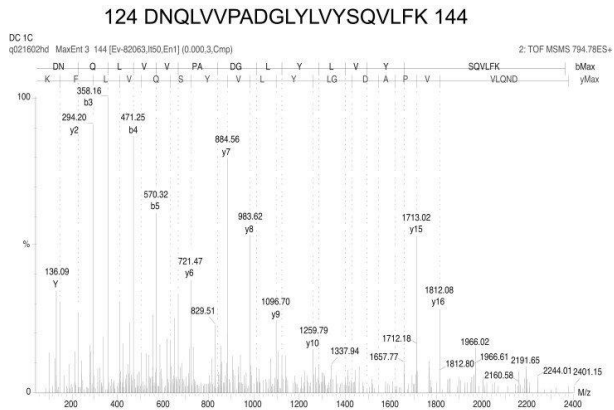

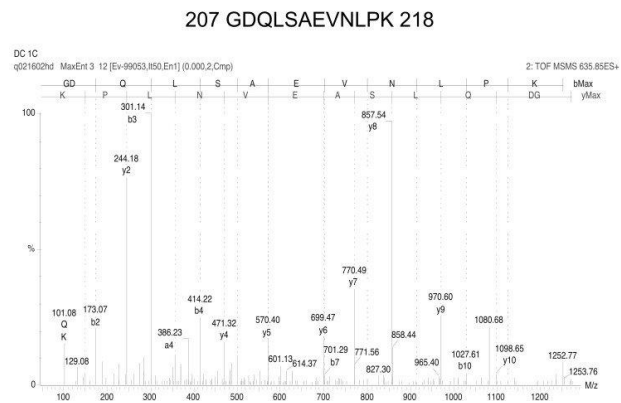

7B

Figure 7: MS/MS spectra of TNF and IL12

Two annotated MS/MS spectra, covering different parts of the corresponding proteins, are shown for each protein, but the identification is secured by MS/MS data obtained on other peptides (see table 2)  Part A: identification of IL12. Part B: identification of TNF

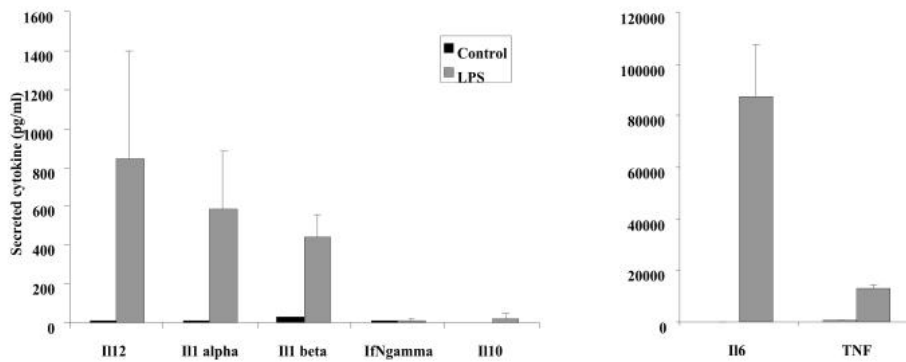

Figure 8: cytokinic profiles of activated dendritic cells

The supernatants produced by dendritic cells and used for the proteomics analysis were also analysed for the expression of various cytokines by specific immunoassays in microtiter plates. The results are expressed in pg/ml. Black bars: expression levels by immature dendritic cells. Grey bars: expression levels by dendritic cells activated for 24 hours by LPS. The results are the mean of 3 independent experiments.

Table1: Proteins identified in the medium conditioned by J774 cells
Short name: name indicated on the figure 3. Name: name indicated in Swiss-Prot. Acc: SwissProt accession number. Mw/pI: theoretical Mw and pI calculated from the sequence. %C: coverage in % of the sequence. Nb pep: number of peptides matching with the predicted protein. Ident: identification method: M= Maldi MS, Q= Qtof MS/MS. Score: Mascot score for the proposed identification

Table 2: Proteins identified in the medium conditioned by dendritic cells
Short name: name indicated on the figure 3. Name: name indicated in Swiss-Prot. Acc: SwissProt accession number. Mw/pI: theoretical Mw and pI calculated from the sequence. %C: coverage in % of the sequence. Nb pep: number of peptides matching with the predicted protein. Ident: identification method: M= Maldi MS, Q= Qtof MS/MS. Score: Mascot score for the proposed identification

Table1: Proteins identified in the medium conditioned by J774 cells

Short name: name indicated on the figure 4. Name: name indicated in Swiss-Prot. Acc: SwissProt accession number. Mw/pI: theoretical Mw and pI calculated from the sequence. %C: coverage in % of the sequence. Nb pep: number of peptides matching with the predicted protein. Ident: identification method: M= Maldi MS, Q= Qtof MS/MS. Score: Mascot score for the proposed identification

| PROTEIN | ACC. | MW / pI | %C | nb pep. | Ident. | Score |
|---|---|---|---|---|---|---|
| MACROPHAGE COLONY SIMULATING FACTOR I RECEPTOR PRECURSOR | P09581 | 109110 / 5,84 | 10% | 8 | M. | 60 |
| MACROPHAGE COLONY SIMULATING FACTOR I RECEPTOR PRECURSOR | P09581 | 109110 / 5,84 | 5% | 5 | M. | 49 |
| GELSOLIN PRECURSOR | P13020 | 85888 / 5,83 | 27% | 16 | M. | 111 |
| BETA-GLUCURONIDASE PRECURSOR | P12265 | 74192 / 6,16 | 28% | 14 | M | 110 |
| MOESIN | P26041 | 67594 / 6,24 | 60% | 58 | M | 432 |
| L-PLASTIN (65KDA MACROPHAGE PROTEIN) | Q61233 | 70105 / 5,20 | 67% | 40 | M | 285 |
| GRANULINS PRECURSOR | P28798 | 63413 / 6,42 | 8% | 5 | M | 56 |
| GLUCOSE REGULATED PROTEIN | Q99LF6 | 56643 / 5,88 | 50% | 28 | M | 233 |
| PYRUVATE KINASE, ISOZYME M2 | P52480 | 57719 / 7,42 | 43% | 23 | M | 212 |
| PYRUVATE KINASE, ISOZYME M2 | P52480 | 57719 / 7,42 | 63% | 35 | M | 300 |
| MACROPHAGE METALLOELASTASE PRECURSOR | P34960 | 53799 / 9,12 | 34% | 20 | M | 214 |
| 6-PHOSPHOGLUCONATE DEHYDROGENASE, DECARBOXYLATING | Q9DCD0 | 53082 / 6,88 | 41% | 18 | M | 167 |
| LYSOSOMAL PROTECTIVE PROTEIN PRECURSOR | P16675 | 53809 / 5,56 | 42% | 25 | M | 223 |
| MONOCYTE DIFFERENTIATION ANTIGEN CD14 PRECURSOR | P10810 | 39179 / 5,08 | 20% | 6 | M | 84 |
| MONOCYTE DIFFERENTIATION ANTIGEN CD14 PRECURSOR | P10810 | 39179 / 5,08 | 18% | 5 | M | 60 |
| BOVIN : ALPHA-2-HS-GLYCOPROTEIN PRECURSOR | P12763 | 38394 / 5,26 | 14% | 5 | M | 47 |
| FRUCTOSE-BISPHOSPHATE ALDOLASE A | P05064 | 39200 / 8,39 | 36% | 13 | M | 90 |
| L-LACTATE DEHYDROGENASE A CHAIN | P06151 | 36344 / 7,77 | 48% | 19 | M | 203 |
| L-LACTATE DEHYDROGENASE A CHAIN | P06151 | 36344 / 7,77 | 29% | 13 | M | 152 |
| PROLIFERATING CELL NUCLEAR ANTIGEN | P17918 | 28766 / 4,66 | 31% | 8 | M | 66 |
| RAT : TROPOMYOSIN | Q63600 | 28703 / 4,69 | 26% | 8 | M | 90 |
| 14-3-3 PROTEIN EPSILON | P62259 | 29155 / 4,63 | 65% | 22 | M | 193 |
| GAG PROTEIN (FRAG. C-TER)   +1282+1702+2295 | Q60588 | 60475 / 7,63 | 23% | 12 | M | 95 |
| PHOSPHOGLYCERATE MUTASE 1 | Q9DBJ1 | 28683 / 6,75 | 64% | 15 | M | 158 |
| PROTEASOME SUBUNIT ALPHA TYPE 7 | Q9Z2U0 | 27838 / 8,59 | 48% | 11 | M | 100 |
| PROTEASOME SUBUNIT ALPHA TYPE 2 | P49722 | 25778 / 8,42 | 54% | 9 | M | 78 |
| HETEROGENEOUS NUCLEAR RIBONUCLEOPROTEIN A1 | P49312 | 34044 / 9,27 | 29% | 10 | M | 124 |
| PROTEASOME SUBUNIT BETA TYPE 1 | O09061 | 26355 / 7,67 | 36% | 9 | M | 106 |
| NUCLEOSIDE DIPHOSPHATE KINASE B | Q01768 | 17352 / 6,97 | 69% | 11 | M | 66 |
| PEPTIDYL-PROLYL CIS-TRANS ISOMERASE A | P17742 | 17829 / 7,88 | 67% | 15 | M | 142 |
| COFILIN, NON-MUSCLE ISOFORM | P18760 | 18417 / 8,26 | 66% | 13 | M | 118 |
| PEPTIDYL-PROLYL CIS-TRANS ISOMERASE A | P17742 | 17829 / 7,88 | 64% | 11 | M | 112 |
| BETA-2-MICROGLOBULIN PRECURSOR | P01887 | 13814 / 7,79 | 50% | 12 | M | 82 |
| MACROPHAGE MIGRATION INHIBITORY FACTOR | P34884 | 12365 / 7,27 | 23% | 4 | M | 49 |
| COMPEMENT C3 PRECURSOR (FRAG. N-TER) | P01027 | 186365 / 6,39 | 28% | 46 | M | 248 |
| MACROPHAGE METALLOELASTASE PRECURSOR | P34960 | 53799 / 9,12 | 50% | 27 | M | 292 |
| LIPOPROTEIN LIPASE PRECURSOR | P11152 | 53093 / 8,20 | 31% | 15 | M | 120 |
| CATHEPSIN B PRECURSOR | P10605 | 37256 / 5,57 | 51% | 20 | M | 141 |
| CATHEPSIN Z PRECURSOR | Q9WUU7 | 33974 / 6,13 | 28% | 8 | M | 71 |
| PEROXIREDOXIN 1 | P35700 | 22162 / 8,26 | 78% | 21 | M | 244 |
| PROTEASOME SUBUNIT ALPHA TYPE 6 | Q9QUM9 | 27355 / 6,34 | 35% | 9 | M | 107 |
| PROTEASOME SUBUNIT BETA TYPE 7 PRECURSOR | P70195 | 29872 / 8,14 | 12% | 7 | M | 52 |

| Abbr. | Protein | ACC. | MW / pI | %C | nb pep | ident | score |
|---|---|---|---|---|---|---|---|
| IL1 RA | INTERLEUKIN-1 RECEPTOR ANTAGONIST PRECURSOR | P25085 | 20261 / 5,82 | 35% | 7 | Q | 207 |
| IL1 RA | INTERLEUKIN-1 RECEPTOR ANTAGONIST PRECURSOR | P25085 | 20261 / 5,82 | 29% | 5 | Q | 168 |
| TNF | TUMOR NECROSIS FACTOR PRECURSOR | P06804 | 25879 / 5,01 | 20% | 5 | M | 43 |
| TNF | TUMOR NECROSIS FACTOR PRECURSOR | P06804 | 25879 / 5,01 | 43% | 8 | Q | 296 |
| IL12 | INTERLEUKIN-12 BETA CHAIN PRECURSOR | P43432 | 38211 / 6,14 | 5% | 2 | Q | 57 |
| IL12 | INTERLEUKIN-12 BETA CHAIN PRECURSOR | P43432 | 38211 / 6,14 | 9% | 3 | Q | 117 |
| Cat-D | CATHEPSIN D PRECURSOR | P18242 | 44925 / 6,71 | 27% | 9 | Q | 333 |
| MMP12 | MACROPHAGE METALLOELASTASE PRECURSOR | P34960 | 53799 / 9,12 | 12% | 7 | Q | 230 |
| C3 | COMPLEMENT C3 PRECURSOR (N-TERMINAL) | P01027 | 186365 / 6,39 | 7% | 10 | Q | 665 |
| C3 | COMPLEMENT C3 PRECURSOR (N-TERMINAL) | P01027 | 186365 / 6,39 | 11% | 16 | Q | 702 |
| C3 | COMPLEMENT C3 PRECURSOR (N-TERMINAL) | P01027 | 186365 / 6,39 | 16% | 24 | M | 191 |
| Compl Fact B | COMPLEMENT FACTOR B PRECURSOR | P04186 | 84951 / 7,18 | 4% | 3 | Q | 91 |
| Endop | BOVIN : ENDOPIN 1 | Q9TTE1 | 46175 / 5,57 | 15% | 4 | Q | 107 |
| CCAP | CYCLOPHILIN C-ASSOCIATED PROTEIN | O35649 | 64014 / 4,91 | 15% | 8 | Q | 245 |
| A2 HS | BOVIN : ALPHA-2-HS GLYCOPROTEIN PRECURSOR | P12763 | 38394 / 5,26 | 47% | 15 | M | 124 |
| cath-Z | CATHEPSIN Z PRECURSOR | Q9WUU7 | 33974 / 6,13 | 23% | 7 | M | 57 |
| NPC2 | EPIDIDYMAL SECRETORY PROTEIN E1 PRECURSOR | Q9Z0J0 | 16432 / 7,59 | 44% | 7 | Q | 155 |
| NPC2 | EPIDIDYMAL SECRETORY PROTEIN E1 PRECURSOR | Q9Z0J0 | 16432 / 7,59 | 14% | 2 | Q | 21 |
| cath-B | CATHEPSIN B PRECURSOR | P10605 | 37256 / 5,57 | 44% | 16 | M | 142 |
| transf | **BOVIN** : SEROTRANSFERRIN PRECURSOR | Q29443 | 77703 / 6,75 | 63% | 41 | M | 185 |
| LCAL2 | LIPOCALIN 2 PRECURSOR | P11672 | 22861 / 8,96 | 52% | 13 | M | 89 |
| Prx1 | PEROXIREDOXIN 1 | P35700 | 22162 / 8,26 | 64% | 17 | M | 143 |

Table 2: Proteins identified in the medium conditioned by dendritic cells
Short name: name indicated on the figure 6. Name: name indicated in Swiss-Prot. Acc: SwissProt accession number. Mw/pI: theoretical Mw and pI calculated from the sequence. %C: coverage in % of the sequence. Nb pep: number of peptides matching with the predicted protein. Ident: identification method: M= Maldi MS, Q= Qtof MS/MS. Score: Mascot score for the proposed identification